\documentstyle[12pt]{article}
\topmargin - 1.5 cm 

\begin{document}

\begin{center}
{\large{\bf SU(2) -- Monopole: Interbasis Expansion}}
\end{center}
\begin{center}
{\bf L.G. Mardoyan${}^1$, A.N. Sissakian${}^2$}
\end{center}
\begin{center}
{\large Joint Institute for Nuclear Research}
\end{center}
\begin{center}
141980, Dubna, Moscow Region, Russia
\end{center}

\footnotetext[1]{E-mail: mardoyan@thsun1.jinr.dubna.su}
\footnotetext[2]{E-mail: sisakian@jinr.dubna.su}

\begin{center}
\underline{Abstract}
\end{center}
{\small This article deals with a nonrelativistic quantum mechanical
study of a charge--dyon system with the $SU(2)$--monopole in five
dimensions. The Schr\"odinger equation for this system is separable
in the hyperspherical and parabolic coordinates. The problem of interbasis
expansion of the wave functions is completely solved. The coefficients
for the expansion of the parabolic basis in terms of the
hyperspherical basis can be expressed through the Clebsch--Gordan
coefficients of the group $SU(2)$.}

\section {Introduction}

A charge--dyon system with the $SU(2)$--monopole in the space
${\rm I \!R}^5$ is described by the equation [1]
\begin{equation}
\frac{1}{2m}\left(-i\hbar \frac{\partial}{\partial x_j}-
\hbar A^a_j \hat T_a\right)^2\psi + \frac{\hbar^2}{2Mr^2}
{\hat T}^2\psi - \frac{e^2}{r}\psi = \epsilon \psi
\end{equation}
where $j=0,1,2,3,4$; $a=1,2,3$. The operators ${\hat T}_a$ are
the generators of the $SU(2)$ group and satisfy the commutation
relations
\begin{eqnarray*}
[\hat T_a,\hat T_b] = i\epsilon_{abc}\hat T_c
\end{eqnarray*}

The triplet of five--dimensional vectors $\vec A^a$ is given
by the expressions
\begin{eqnarray*}
\vec A^1 &=& \frac{1}{r(r + x_0)}(0, -x_4, -x_3, x_2, x_1)  \\ [2mm]
\vec A^2 &=& \frac{1}{r(r + x_0)}(0, x_3, -x_4, -x_1, x_2)  \\ [2mm]
\vec A^3 &=& \frac{1}{r(r + x_0)}(0, x_2, -x_1, x_4, -x_3)
\end{eqnarray*}

Every term of the triplet $A_j^a$ coincides with the vector potential of
the 5D Dirac monopole [2] with a unit topological charge and the line of
singularity along the nonpositive part of the $x_0$--axis. The vectors
$A_j^a$ are orthogonal to each other
\begin{eqnarray*}
A^a_jA^b_j = \frac{1}{r^2}\frac{(r - x_0)}{(r + x_0)}
{\delta}_{ab}
\end{eqnarray*}
and also to the vector $\vec x=(x_0,x_1,x_2,x_3,x_4)$.

The eigenvalues of the energy $(N=0,1,2...)$
\begin{eqnarray}
{\epsilon}_N^T=-\frac{me^4}{2\hbar^2(\frac{N}{2}+2)^2}
\end{eqnarray}
for fixed $T$ are degenerated with multiplicity [3]
\begin{eqnarray*}
g_N^T =\frac{1}{12}(2T+1)^2\left(\frac{N}{2}-T+1\right)
\left(\frac{N}{2}-T+2\right) \\ [3mm]
\left\{\left(\frac{N}{2}-T+2\right)
\left(\frac{N}{2}-T+3\right)+2T(N+5)\right\}
\end{eqnarray*}
For $T=0$ and $N=2n$ (even) the r.h.s. of the last formula is equal
to $(n+1)(n+2)^2(n+3)/12$, i.e., to the degeneracy of pure Coulomb
levels.

The article is organized as follows: In Section 2, we describe the
hyperspherical and parabolic bases for the charge--dyon system
with the $SU(2)$--monopole in a way adapted to the introduction of
interbasis expansions. In Section 3, we prove an additional orthogonality
property for hyperspherical radial wave functions of the given
hypermomentum $\lambda$. In Section 4, by using the property of
biorthogonality of the hyperspherical basis, we calculate the coefficients
of interbasis expansion between hyperspherical and parabolic bases.

\section {Hyperspherical and parabolic bases}

The variables in Eq.(1) are separated in the hyperspherical and
parabolic coordinates.

Let us introduce in ${\rm I \!R}^5$ the hyperspherical coordinates
$r\in [0,\infty)$, $\theta \in [0,\pi]$, $\alpha \in [0,2\pi)$,
$\beta \in [0,\pi]$, $\gamma \in [0,4\pi)$ according to
\begin{eqnarray*}
x_0 &=& r\cos \theta  \nonumber \\ [2mm]
x_1 + ix_2 &=& r \sin \theta \cos \frac{\beta}{2}e^{i\frac{\alpha
+\gamma}{2}}  \\ [2mm]
x_3 + ix_4 &=& r \sin \theta \sin \frac{\beta}{2}e^{i\frac{\alpha -
\gamma}{2}}  \nonumber
\end{eqnarray*}
Since
\begin{eqnarray*}
iA_j^a\frac{\partial}{\partial x_j}=\frac{2}{r(r+x_0)}{\hat L}_a,
\end{eqnarray*}
where
\begin{eqnarray*}
\hat L_1 &=& \frac{i}{2}\left[D_{41}(x)+D_{32}(x)\right]  \\ [2mm]
\hat L_2 &=& \frac{i}{2}\left[D_{13}(x)+D_{42}(x)\right]  \\  [2mm]
\hat L_3 &=& \frac{i}{2}\left[D_{12}(x)+D_{34}(x)\right]
\end{eqnarray*}
and
\begin{eqnarray*}
D_{ij}(x) = - x_i\frac{\partial}{\partial x_j} +
x_j\frac{\partial}{\partial x_i}
\end{eqnarray*}
Eq.(1) in the hyperspherical coordinates assumes the form
\begin{equation}
\left(\Delta_{r \theta}
- \frac{{\hat L}^2}{r^2 \sin^2 \theta/2}
- \frac{{\hat J}^2}{r^2 \cos^2 \theta/2}\right)\psi +
\frac{2m}{\hbar^2}\left(\epsilon + \frac{e^2}{r}\right)\psi=0
\end{equation}
Here
\begin{eqnarray*}
\Delta_{r \theta} =
\frac{1}{r^4}\frac{\partial}{\partial r}
\left(r^4 \frac{\partial}{\partial r}\right) +
\frac{1}{r^2 \sin^3 \theta}\frac{\partial}{\partial \theta}
\left(\sin^3 \theta \frac{\partial}{\partial \theta}\right)
\end{eqnarray*}
and ${\hat J}_a = {\hat L}_a + {\hat T}_a$. Emphasize that
\begin{eqnarray*}
[\hat L_a,\hat L_b] = i\epsilon_{abc}\hat L_c,\,\,\,\,\,\,
[\hat J_a,\hat J_b] = i\epsilon_{abc}\hat J_c
\end{eqnarray*}
The solution of Eq.(3) is of the form [3]
\begin{eqnarray}
{\psi}^{sph}=R_{N \lambda}(r)Z_{\lambda LJ}(\theta)
G_{LTm't'}^{JM}(\alpha,\beta,\gamma;\alpha_T,\beta_T,\gamma_T)
\end{eqnarray}
where $G$ are the eigenfunctions of ${\hat L}^2$, ${\hat T}^2$ and
${\hat J}^2$ with the eigenvalues $L(L+1)$, $T(T+1)$ and $J(J+1)$;
${\alpha}_T$, ${\beta}_T$ and ${\gamma}_T$ are the coordinates of the
space group of $SU(2)$ and have the form
\begin{eqnarray*}
G = \sqrt{\frac{(2L+1)(2T+1)}{4{\pi}^4}}
\sum_{M=m+t}\left(JM|L,m';T,t'\right)D_{mm'}^L(\alpha,\beta,\gamma)
D_{tt'}^T(\alpha_T,\beta_T,\gamma_T)
\end{eqnarray*}
Here $\left(JM|L,m';T,t'\right)$ are the Clebsh--Gordan coefficients,
and $D_{mm'}^L$ and $D_{tt'}^T$ are the Wigner functions.

The functions $Z_{\lambda LJ}(\theta)$ and $R_{N \lambda}(r)$
normalized by the conditions
\begin{eqnarray*}
\int \limits_{0}^{\pi} {\sin}^3 \theta
Z_{{\lambda}'LJ}(\theta)Z_{\lambda LJ}(\theta)d\theta =
\delta_{{\lambda}'{\lambda}}
\end{eqnarray*}
\begin{eqnarray}
\int \limits_{0}^{\infty} r^4 R_{N' \lambda}(r)
R_{N \lambda}(r)dr = \delta_{N'N}
\end{eqnarray}
are given by the formulae
\begin{eqnarray}
Z_{\lambda LJ}(\theta) &=& N_{JLT}^{\lambda}
(1-\cos \theta)^L(1+\cos \theta)^J
P_{\lambda-L-J}^{(2L+1,2J+1)}(\cos \theta) \\ [2mm]
R_{N \lambda}(r) &=& C_{N \lambda}e^{-\kappa r}(2\kappa r)^\lambda
F\left(-\frac{N}{2}+\lambda; 2\lambda+4; 2\kappa r\right)
\end{eqnarray}
Here $P_n^{(\alpha,\beta)}(x)$ are the Jacobi polynomials;
$\kappa = 2/r_0(N+4)$, $r_0={\hbar}^2/me^2$ is the Bohr radius.
The normalization constants $N_{JLT}^{\lambda}$ and $C_{N \lambda}$
equal
\begin{eqnarray*}
C_{LJT}^\lambda &=& \sqrt{\frac
{(2\lambda+3)(\lambda-J-L)!\Gamma(\lambda+J+L+3)}
{2^{2J+2L+3}\Gamma(\lambda+J-L+2)\Gamma(\lambda-J+L+2)}} \\ [2mm]
C_{N \lambda} &=& \frac{32}{(N + 4)^3}\frac{1}{(2\lambda+3)!}
\sqrt{\frac{(\frac{N}{2}+\lambda+3)!}{r_0^5(\frac{N}{2}-\lambda)!}}
\end{eqnarray*}
The quantum numbers run over the values $|L-T|\leq J\leq L+T$;
$\lambda=L+J,L+J+1,...,N/2.$

In the parabolic coordinates
\begin{eqnarray*}
x_0 &=& \frac{1}{2}\left(\xi - \eta\right)   \\ [2mm]
x_1 + ix_2 &=& \sqrt{\xi \eta} \cos \frac{\beta}{2}e^{i\frac{\alpha
+\gamma}{2}}  \\ [2mm]
x_3 + ix_4 &=& \sqrt{\xi \eta} \sin \frac{\beta}{2}e^{i\frac{\alpha -
\gamma}{2}}
\end{eqnarray*}
where $\xi, \eta \in [0,\infty)$, upon the substitution
\begin{eqnarray*}
{\psi}^{par}=f_1(\xi)f_2(\eta)
G_{LTm't'}^{JM}(\alpha,\beta,\gamma;\alpha_T,\beta_T,\gamma_T)
\end{eqnarray*}
the variables in Eq.(1) are separated, which results in the
system of equations
\begin{eqnarray*}
\frac{1}{\xi}\frac{d}{d \xi}\left({\xi}^2 \frac{df_1}{d \xi}\right) +
\left[\frac{m\epsilon}{2{\hbar}^2} - \frac{1}{\xi}J(J+1)
+\beta_1\right]f_1 = 0  \\ [2mm]
\frac{1}{\eta}\frac{d}{d \eta}\left({\eta}^2 \frac{df_2}{d \eta}\right) +
\left[\frac{m\epsilon}{2{\hbar}^2} - \frac{1}{\eta}L(L+1)
+\beta_2\right]f_2 = 0
\end{eqnarray*}
where
\begin{eqnarray}
\beta_1 + \beta_2 = \frac{me^2}{{\hbar}^2}
\end{eqnarray}
At $T=0$ (i.e. $J=L$), these equations coincide with the equations for a
five dimensional Coulomb problem in the parabolic coordinate [4],
and consequently,
\begin{eqnarray}
\psi^{par} = {\kappa}^3\sqrt{2r_0}f_{n_1J}(\xi)f_{n_2L}(\eta)
G_{LTm't'}^{JM}(\alpha,\beta,\gamma;\alpha_T,\beta_T,\gamma_T)
\end{eqnarray}
where
\begin{eqnarray*}
f_{pq}(x) = \frac{1}{(2q+1)!}\sqrt{\frac{(p+2q+1)!}{p!}}
\exp{\left(-\frac{\kappa x}{2}\right)}(\kappa x)^q
F\left(-p; 2q+2; \kappa x\right)
\end{eqnarray*}
Here $n_1$ and $n_2$ are non-negative integers
\begin{eqnarray*}
n_1 = -J-1+\frac{\beta_1}{\kappa},\,\,\,\,\,\,\,
n_2 = -L-1+\frac{\beta_2}{\kappa}
\end{eqnarray*}
from which and (2), (8) it follows that the parabolic quantum numbers
$n_1$, $n_2$, $J$ and $L$ are connected with the principal quantum
number $N$ as follows:
\begin{eqnarray*}
N = 2(n_1+n_2+J+L)
\end{eqnarray*}

\section {Biorthogonality of the radial wave functions}

We shall prove that along with the condition (5) the radial wave
functions $R_{N \lambda}(r)$ satisfy the following "additional"
orthogonality condition:
\begin{eqnarray}
J_{\lambda \lambda'} = \int \limits_{0}^{\infty} r^2 R_{N \lambda'}(r)
R_{N \lambda}(r)dr = \frac{16}{r_0^2(N+4)^2}\frac{1}{2\lambda+3}
\delta_{\lambda \lambda'}
\end{eqnarray}
This new relation shall prove useful when dealing with the interbasis
expansions in the next Section. The proof of the formula (10) is as
follows.

In the integral appearing in (10), we replace $R_{N \lambda}(r)$ and
$R_{N \lambda'}(r)$ by their expressions (7). Then, we take the confluent
hypergeometric function in (7) as an finite sum
\begin{eqnarray*}
F\left(-\frac{N}{2}+\lambda; 2\lambda+4; 2\kappa r\right) =
\sum_{s=0}^{\frac{N}{2}-\lambda}
\frac{\left(-\frac{N}{2}+\lambda\right)_s}{(2\lambda+4)_s}
\frac{(2\kappa r)^s}{s!}
\end{eqnarray*}
and perform the integration term by term with the help of the
formula [5]
\begin{eqnarray}
\int \limits_{0}^{\infty} e^{-\lambda x} x^\nu
F(\alpha, \gamma; kx) \, dx =
\frac{\Gamma(\nu+1)}{\lambda^{\nu+1}} \,
{_2F}_1 \left( \alpha, \nu+1, \gamma ; \frac{k}{\lambda} \right).
\end{eqnarray}
By using
\begin{eqnarray}
{_2F}_1 \left(a, b; c; 1\right) = \frac{\Gamma(c)\Gamma(c-a-b)}
{\Gamma(c-a)\Gamma(c-b)}
\end{eqnarray}
we arrive at
\begin{eqnarray}
J_{\lambda \lambda'} = \frac{16}{r_0^2(N+4)^2}
\frac{\Gamma(\lambda+\lambda'+3)}{(2\lambda+3)}\sqrt{\frac
{\left(\frac{N}{2}+\lambda+3\right)!}{\left(\frac{N}{2}-\lambda\right)!
\left(\frac{N}{2}-\lambda'\right)!
\left(\frac{N}{2}+\lambda'+3\right)!}} \nonumber \\ [2mm]
\sum_{s=0}^{\frac{N}{2}-\lambda}
\frac{\left(-\frac{N}{2}+\lambda\right)_s}{s!}
\frac{(\lambda+\lambda'+3)_s}{(2\lambda+4)_s}
\frac{\Gamma\left(\frac{N}{2}-\lambda - s +1\right)}
{\Gamma(\lambda - \lambda'- s +1)}
\end{eqnarray}
By introducing formula [6]
\begin{eqnarray}
\frac{\Gamma(z)}{\Gamma(z-n)} =
(-1)^n\frac{\Gamma(-z+n+1)}{\Gamma(-z+1)}
\end{eqnarray}
into (13), the sum over $s$ can be expressed in terms of the
${_2F}_1$ Gauss hypergeometric function of argument 1. We thus obtain
\begin{eqnarray}
J_{\lambda \lambda'} = \frac{16}{r_0^2(N+4)^2}
\frac{1}{\lambda+\lambda'+3}\sqrt{\frac
{\left(\frac{N}{2}-\lambda\right)!\left(\frac{N}{2}+\lambda+3\right)!}
{\left(\frac{N}{2}-\lambda'\right)!
\left(\frac{N}{2}+\lambda'+3\right)!}} \nonumber \\ [2mm]
\frac{1}{\Gamma(\lambda - \lambda' + 1)\Gamma(\lambda' - \lambda + 1)}
\end{eqnarray}
Equation (10) then easily follows from (15) since
$[\Gamma(\lambda - \lambda' + 1)\Gamma(\lambda' - \lambda + 1)]^{-1}=
\delta_{\lambda \lambda'}$.

The result provided by formula (10) generalizes the one for the hydrogen
atom [6]. Such unusual orthogonality properties are connected with the
accidental degeneracies of the energy spectrum for the charge-dyon
system with the $SU(2)$-monopole.

\section {Interbasis expansion}

The connection between hyperspherical $(r, \theta, \alpha, \beta, \gamma)$
and parabolic $(\xi, \eta, \alpha, \beta, \gamma)$ coordinates is
\begin{eqnarray}
\xi = r(1+\cos \theta),\,\,\,\,\,\,\,\, \eta = r(1-\cos \theta)
\end{eqnarray}

Now, we can write, for fixed value energy, the parabolic bound states (9)
as a coherent quantum mixture of the hyperspherical bound states (4)
\begin{eqnarray}
\psi^{par}=\sum_{\lambda=T}^{N/2} W_{n_1 n_2 JL}^\lambda\psi^{sph}
\end{eqnarray}
By virtue of Eq.(16), the left-hand side of (17) can be rewritten in
hyperspherical coordinates. Then, by substituting $\theta=0$ in the
so-obtained equation and by taking into account that
\begin{eqnarray*}
P_n^{(\alpha,\beta)}(1) = \frac{(\alpha +1)_n}{n!}
\end{eqnarray*}
we get an equation that depends only on the variable $r$. Thus, we can use
the orthogonality relation (10) on the hypermomentum quantum numbers
$\lambda$. This yields
\begin{eqnarray}
W_{n_1 n_2 JL}^\lambda = \frac{1}{(2J+1)!(2\lambda+3)!}
E_{\lambda JL}^{n_1n_2}K_{\lambda JL}^{nn_1}
\end{eqnarray}
where
\begin{eqnarray}
E_{\lambda JL}^{n_1n_2} =
\sqrt{(2\lambda+3)(\lambda-J-L)!\left(\frac{N}{2}+\lambda+3\right)!} \nonumber \\ [2mm]
\left[\frac{\Gamma(\lambda+J-L+2)(n_1+2J+1)!(n_2+2L+1)!}
{(n_1)!(n_2)!\left(\frac{N}{2}-\lambda\right)!
\Gamma(\lambda-J+L+2)\Gamma(\lambda+J+L+3)}\right]^{1/2}
\end{eqnarray}
\begin{eqnarray}
K_{\lambda JL}^{nn_1}=\int \limits_{0}^{\infty} e^{-x}
x^{\lambda+J+L+2}F(-n_1, 2J+2; x)
F\left(-\frac{N}{2}+\lambda, 2\lambda+4; x\right)dx
\end{eqnarray}
To calculate the integral $K_{\lambda JL}^{nn_1}$, it is sufficient to write
the confluent hypergeometric function $F(-n_1, 2J+2; x)$ as a series,
integrate according to (11) and use the formula (12) for the
summation of the hypergeometric function ${_2F}_1$. We thus obtain
\begin{eqnarray}
K_{\lambda JL}^{nn_1}=\frac{(2\lambda+3)!
\left(\frac{N}{2}-J-L\right)!\Gamma(\lambda+J+L+3)}
{(\lambda-J-L)!\left(\frac{N}{2}+\lambda+3\right)!} \nonumber \\ [2mm]
{_3F}_2 \left\{
\begin{array}{l}
-n_1,  -\lambda+J+L, \lambda+J+L+3 \\
2J+2,   -\frac{N}{2}+J+L
\end{array}
\biggr| 1 \right\}
\end{eqnarray}
The introduction of (19) and (21) into (18) gives
\begin{eqnarray}
W_{n_1 n_2 JL}^\lambda =
\left[\frac{(2\lambda+3)\Gamma(\lambda+J+L+3)
(n_1+2J+1)!(n_2+2L+1)!}{(n_1)!(n_2)!(\lambda-J-L)!
\left(\frac{N}{2}-\lambda\right)!\left(\frac{N}{2}+\lambda+3\right)!}
\right]^{1/2}\nonumber \\ [2mm]
\frac{\left(\frac{N}{2}-J-L\right)!}{(2J+1)!}
\sqrt{\frac{\Gamma(\lambda+J-L+2)}{\Gamma(\lambda-J+L+2)}}
{_3F}_2 \left\{
\begin{array}{l}
-n_1,  -\lambda+J+L, \lambda+J+L+3 \\
2J+2,   -\frac{N}{2}+J+L
\end{array}
\biggr| 1 \right\}
\end{eqnarray}

The next step is to show that the interbasis coefficients (22) are the
Clebsch-Gordan coefficients for the group $SU(2)$. It is known that
the Clebsch-Gordan coefficient can be written as [7]
\begin{eqnarray}
C_{a \alpha ; b \beta}^{c\gamma} = (-1)^{a-\alpha}
\delta_{\gamma,\alpha+\beta}
\frac{(a+b-\gamma)!(b+c-\alpha)!}{\sqrt{(b-\beta)!(b+\beta)!}}
\nonumber \\ [2mm]
\left[ \frac{(2c+1) (a+\alpha)! (c+\gamma)!}
{(a-\alpha)!(c-\gamma)!(a+b+c+1)!(a+b-c)!(a-b+c)!
(b-a+c)!} \right] ^{1/2} \nonumber \\ [2mm]
{_3F}_2
\left\{
\begin{array}{l}
-a-b-c-1, -a+\alpha, -c+\gamma \\
-a-b+\gamma,  -b-c+\alpha  \\
\end{array}
\biggr| 1 \right\}
\end{eqnarray}
By using the formula [8]
\begin{eqnarray*}
{_3F}_2 \left\{
\begin{array}{l}
s, s', -N \\
t', 1-N-t  \\
\end{array}
\biggr| 1 \right\} =
\frac{(t+s)_N}{(t)_N}
\> {_3F}_2 \left\{
\begin{array}{l}
s, t' - s', -N \\
t', t+s  \\
\end{array}
\biggr| 1 \right\}
\end{eqnarray*}
equation (23) can be rewritten in the form
\begin{eqnarray}
C_{a \alpha ; b \beta}^{c\gamma} =
\left[ \frac{(2c+1)(b-a+c)!(a+\alpha)!(b+\beta)!(c+\gamma)!}
{(b-\beta)!(c-\gamma) ! (a+b-c)!(a-b+c)!(a+b+c+1)!} \right] ^{1/2}
\nonumber \\ [2mm]
\delta_{\gamma,\alpha+\beta}\frac{(-1)^{a-\alpha}}{\sqrt{(a-\alpha)!}}
\frac{(a+b-\gamma)!}{(b-a+\gamma)!}
{_3F}_2 \left\{
\begin{array}{l}
-a+\alpha, c+\gamma+1, -c+\gamma \\
\gamma-a-b,  b-a+\gamma+1  \\
\end{array}
\biggr| 1 \right\}
\end{eqnarray}
By comparing (24) and (22), we finally obtain the desired
representation
\begin{eqnarray}
W_{n_1 n_2 JL}^\lambda = (-1)^{n_1}
C_{\frac{N-2J+2L+2}{4},L+\frac{n_2-n_1+1}{2};
\frac{N+2J-2L+2}{4},J+\frac{n_1-n_2+1}{2}}^{\lambda+1,J+L+1}
\end{eqnarray}
At $T=0$ (i.e. $J=L$) formula (25) turns into the formula for the
five-dimensional Coulomb problem [4], as would be expected.

\end{document}